\documentstyle[12pt]{article}

\newcommand{\bb}{\begin{equation}}
\newcommand{\ee}{\end{equation}}
\newcommand{\ba}{\begin{eqnarray}}
\newcommand{\ea}{\end{eqnarray}}
\newcommand{\baa}{\begin{eqnarray*}}
\newcommand{\eaa}{\end{eqnarray*}}
\newcommand{\bebib}{\begin{thebibliography}}
\newcommand{\eb}{\end{thebibliography}}
\newcommand{\ci}[1]{\cite{#1}}
\newcommand{\bi}[1]{\bibitem{#1}}
\newcommand{\lab}[1]{\label{#1}}
\newcommand{\re}[1]{(\ref{#1})}

\newcommand\fac[2]{\mbox{$\frac{#1}{#2}$}}

\begin{document}

\begin{titlepage}
\begin{flushright}
  UdeM-LPN-TH-134  \\
   January 1993
\end{flushright}

\vspace*{15mm}

\begin{center}

{\Large \bf Self-Similar Potentials and the \\[3mm]
$\bf q$-Oscillator Algebra at Roots of Unity%
\footnote{To appear in shortened form in Lett.Math.Phys. {\bf 27} (1993)}
}
\vspace{15mm}

{\rm \large   Sergei Skorik%
\footnote{On leave from the Moscow Institute of Physics and
Technology, Moscow, Russia}
}
\medskip

{\em Department of Physics, University of Southern California, \\
Los Angeles, CA 90089-0484, USA
}
\vspace{6mm}

{\rm \large  Vyacheslav Spiridonov%
\footnote{On leave from the Institute for Nuclear Research,
Moscow, Russia}$^,$\footnote{International NSERC Fellow}
}

\medskip

{\em Laboratoire de Physique Nucl\' eaire,
Universit\' e de Montr\' eal, \par
C.P. 6128, succ. A, Montr\' eal, Qu\' ebec, H3C 3J7, Canada}

\end{center}

\vspace*{6mm}
\begin{abstract}

Properties of the simplest class of self-similar potentials are analyzed.
Wave functions of the corresponding Schr\"odinger equation provide
bases of representations of the $q$-deformed Heisenberg-Weyl algebra.
When the parameter $q$ is a root of unity
the functional form of the potentials can be found explicitly.
The general $q^3=1$ and the particular
$q^4=1$ potentials are given by the equianharmonic and (pseudo)lemniscatic
Weierstrass functions respectively.

\medskip


\end{abstract}
\end{titlepage}

\newpage

\section{Introduction}

 The one-dimensional Schr\"odinger equation
\bb
L \psi(x)\equiv -\psi^{\prime\prime}(x)+u(x)\psi(x)=\lambda \psi(x)
\lab{dc1}
\ee
plays an important role in the theory of nonlinear evolution equations,
where it serves as an auxiliary spectral problem allowing
to integrate the KdV-equation. From this point of view
 the KdV-evolution of $u(x)$
in time yields the isospectral deformations of (\ref{dc1}).
In the quantum mechanical context, solutions of (\ref{dc1}) subjected to some
boundary conditions describe physical states of a particle moving on the
line.  In both applications a special attention is
drawn to the class of exactly solvable potentials $u(x)$
whose discrete spectra are given
by some known functions of a quantum number. The powerful approach
to derivation of such potentials is provided by the
factorization method \cite{infeld} based on the Darboux transformations.
In this method one has a set of operators
$L_j=-d^2/dx^2 +u_j(x)$ which are factorized
as products of linear differential operators
$A_j^+=-d/dx +f_j(x), \, A_j^-=d/dx +f_j(x)$ up to some constants
$\lambda_j$: $L_j=A_j^+A_j^-+\lambda_j.$
Since the spectra of the operators $A_j^+A_j^-$ and
$A_j^-A_j^+$ may differ only by one lowest level
(provided $f_j$ has no severe singularities),
the Hamiltonians $L_{j+1}$ and $L_j$ will have close spectral properties
if we set $L_{j+1}\equiv A_{j+1}^+A_{j+1}^-+\lambda_{j+1}=
A_j^-A_j^+ +\lambda_j$. This constraint is equivalent to the following
differential equation relating potentials $u_j$ and $u_{j+1}$:
\bb
f_j^\prime(x)+f_{j+1}^\prime(x)+f_j^2(x)-f^2_{j+1}(x)=\lambda_{j+1}-\lambda_j.
\lab{dds} \ee
The chain
of such equations is called {\it the dressing chain} in the theory of solitons.
It is quite useful for studying symmetries
of the Schr\"odinger equation.
For particular solutions of the dressing chain
 parameters $\lambda_j$ coincide with ordered discrete spectra of the
corresponding Hamiltonians $L_j$.
So, the spectra of harmonic oscillator, Coulomb, and other ``old"
solvable potentials are easily found after the plugging in
(\ref{dds}) the ansatz: $f_j(x)=a(x)j+b(x)+c(x)/j$ \ci{infeld}.

In this Letter we discuss the particular class of the {\it self-similar}
potentials
related to each other by the scaling of argument: $u_{j+1}(x)=q^2u_j(qx)$,
which emerges as a result of the constraints
$f_{j+1}(x)=qf_j(qx),\, \lambda_{j+1}=q^{2}\lambda_j$ imposed on (\ref{dds}).
This class has been studied by A.Shabat \cite{sh} for the
real $x$ and $q$, $0<q<1$,
when it corresponds to ``infinite-soliton" systems, i.e. when the
potentials are reflectionless and decrease slowly at  space
infinities. The corresponding discrete spectra are purely exponential;
they are generated by the $q$-deformed Heisenberg-Weyl, or $q$-oscillator
algebra  \cite{sp1}. In the two limiting cases, $q\to 0$ and $q\to 1$,
such potentials are reduced to the one-soliton and harmonic oscillator
potentials respectively.
In the present paper we consider this self-similar system in the complex domain
of the coordinate $x$ and parameter $q$. First we analyze
the existence and uniqueness of the corresponding functions $f_j(x)$.
Then we characterize the qualitative
structure of singularities exhibited by $f_j(x)$.
When $q$ is a primitive $n$-th root of unity the situation is simplified.
The odd $n$ cases are shown to be related to the finite-gap  potentials.
The even $n$ cases may contain a functional non-uniqueness.
The general $q^3=1$ solution and a special $q^4=1$  solution
are expressed through the equianharmonic and (pseudo)lemniscatic
Weierstrass functions respectively. Algebraically, all
these root-of-unity potentials are naturally related to the
representations of the  $q$-oscillator algebra at $q^n=1$. A more wide
class of the self-similar potentials leading to more complicated
$q$-deformed algebras has been described in \cite{sp2}.
Though some of our results may be  reformulated easily for the latter systems,
they will not be discussed here.

Appearance of the $q$-analogs of the harmonic oscillator and other
 spectrum generating algebras raises special interest
to the self-similar potentials. This is inspired by the recent intensive
discussion of quantum algebras, or $q$-deformations of
Lie algebras \cite {qa}, which bear universal character due to the large number
of applications. The $q$-oscillators
themselves were reinvented as the related objects \cite{qosc}.
It was found that a natural group-theoretical
setting for the basic, or $q$-hypergeometric functions \cite{qsf}
and corresponding orthogonal polynomials is provided by the quantum
algebras when they are realized with the help of
purely finite-difference operators \cite{qpol}.  However,
the $q$-hypergeometric functions are defined in general only at $|q|<1$ and we
were not able to find in the literature an example of
$q$-special function for $q^n=1$ -- the cases when the representation theory
of quantum algebras essentially differs from that for the standard
Lie algebras \cite{root}. The self-similar potentials provide
a realization of the creation and annihilation
operators of the $q$-oscillator algebra in terms of both
differential and finite-difference operators. It is the presence of
the differential part that principally differs our approach
from the mentioned above and leads to the well-defined $q$-special
functions at $q^n=1$.

The paper is organized as follows. In Sect. 2 we describe  the
needed facts about the periodic dressing chain. In Sect. 3 we present
the class of self-similar potentials and discuss its analytical properties.
The central result here is the theorem on the existence
and uniqueness of solutions for the basic differential equation
with deviating argument (\ref{rs}) when $|q|<1$, or $q^n=1$.
The $q^3=1$ and $q^4=1$ cases are considered in detail in Sects. 4 and 5
respectively.  Sect. 6 is devoted to the
description of $q$-oscillator algebra representations
related to the taken $q$-transcendent.  In the Appendix  we present exact
solution of the spectral problem associated with the $q^3=1$ system.

The results obtained in this paper were partially presented at the
Canadian Mathematical Society Meeting (Montr\'eal, December 1992).

\section{The Periodic Dressing Chain}

We discuss here some results  of the  theory of nonlinear evolution
equations on the basis of the dressing chain.

Following \cite{shyam}, one can
rewrite \re{dc1} as a matrix equation and introduce
a chain of associated spectral problems
\bb
\Psi^\prime_j(x)=U_j\Psi_j(x),\qquad
U_j=\left[\matrix{ 0 & 1\cr u_j-\lambda& 0 \cr} \right] , \qquad
\Psi_j=\left(\matrix{ \psi_j\cr \psi^\prime_j\cr}\right) , \quad j\in Z
\lab{dc2}
\ee
solutions of which are related to each other by the transformations
\bb
\Psi_{j+1}(x)=B_j\Psi_j(x),
\lab{dc3}
\ee
where $B_j$ is a $2\times 2$ matrix whose entries are polynomials of
the spectral parameter $\lambda$. The compatibility condition
for (\ref{dc2})-(\ref{dc3}) leads to
the equation
\bb
B_j^\prime=U_{j+1}B_j-B_jU_j.
\lab{dc4}
\ee
When the  map $\psi_j\to \psi_{j+1}$  does not depend on $\lambda$ the
solution of (\ref{dc4}) looks as follows
\bb
B_j=\left[\matrix{ f_j(x)&1\cr f_j^2(x)+\lambda_j-\lambda&f_j(x)\cr}\right],
\lab{dc5}
\ee
\bb
u_j(x)=f_j^2-f_j^\prime +\lambda_j,\qquad u_{j+1}-u_j=2f_j^\prime,
\lab{dc6}
\ee
where $\lambda_j$ are some constants of integration. Substituting the first of
relations \re{dc6} into the second one we get the dressing chain \re{dds}.

Denote by $A_j=B_{j+N-1}\dots B_{j+1} B_j$ the product of Darboux
transformation matrices \re{dc5}. From \re{dc4} it follows that
\bb
A^\prime_j=U_{j+N}A_j-A_jU_j.
\lab{dc8}
\ee
This equation allows to integrate the chain \re{dds} for the special
class of potentials defined by the periodicity condition
\bb
U_{j+N}=U_j, \quad or \qquad f_{j+N}(x)=f_j(x), \qquad
\lambda_{j+N}=\lambda_j,
\lab{dc9}
\ee
which describes a finite-dimensional dynamical system.
For the odd $N$ this leads to the finite-gap potentials \ci{shyam}.

For convenience we fix the index $j$ in \re{dc8}, \re{dc9}, $j=0$, and
set $L\equiv L_0,\, U\equiv U_0,\, A\equiv A_0$.
By its definition the matrix $A$ maps solutions
of the corresponding Schr\"odinger equation onto each other and, hence,
satisfies
the matrix Lax equation $A^\prime=[U,A]$. From the latter, one can derive
the following relation for the  matrix element $a_{12}\equiv \beta(x)$:
\bb \fac12\beta \beta''+(\lambda-u_0)\beta^2-\fac14{\beta'}^2=
det\,A - \fac14(Tr\,A)^2. \label{sp}\ee
Both    $Tr\, A$ and $det\, A$ are easily seen to be integrals of motion; at
$N=2k+1$ one has
$$ det\,A=\prod_{i=0}^{2k} (\lambda-\lambda_i)\,\, , \,\,\,\,\,\,\,\,\, Tr\,A=
(-1)^k (I_0\lambda^k+I_1\lambda^{k-1}+\cdots +I_k).$$
Equation (\ref{sp})  can be rewritten now in the form
\bb u_0-\lambda=F^2-F^\prime,\qquad F(x)\equiv\frac{\pm w-\beta'}{2\beta},
\label{spec}\ee
where
$$w^2=(TrA)^2 - 4detA\equiv -4\prod_{i=0}^{2k}(\lambda-E_i).$$
If all the constants $E_i$ are real and mutually different
then the spectrum of non-singular
potentials $u_j(x)$ has zonal structure
and $E_i$ coincide with the boundaries of gaps. It is well-known that such
potentials can be expressed in terms of the Riemann $\Theta$-function
\cite{its}:
\bb
u(x)=-2\frac{d^2}{dx^2}\ln \Theta(\vec l(x))+const,
\lab{its} \ee
$$
\quad \Theta(\vec l)=\sum_{m_1,\dots,m_k \in Z}
\exp \{\sum_{p,s=1}^k m_p b_{ps} m_s + \sum_{p=1}^k l_p m_p\},
$$
where $l_p(x)$ are linear functions of $x$ whose coefficients together
with the parameters $b_{ps}$ are determined by $E_i$.

We shall not go further in the presentation of general results. For a
detailed account of integrability of the dressing chain at even
period $N$ and in the
case of a more general than (\ref{dc9}) closure condition $u_{j+N}=u_j+\alpha$,
where $\alpha$ is a constant, we refer to the recent work \cite{vs}.

\section{Self-Similar Potentials}

The particular solution for the dynamical system \re{dds}, which we investigate
in the present paper,
is fixed by the following self-similarity condition
\bb
f_{j}(x)=q^jf(q^{j}x), \qquad \mu_j=q^{2j}\mu,\qquad
\mu_j\equiv \lambda_{j+1}-\lambda_j.
\lab{ss1}
\ee
In this case an infinite number of relations \re{dds} is reduced to one
differential equation with the deviating argument which was introduced by
A.Shabat in \cite{sh} %
\footnote{According to the classification of equations with a deviating
 argument \cite{EN}, equation (\ref{rs}) is of the neutral type.} :
\bb {d\over dx}\left(f(x)+qf(qx)\right)+f^2(x)- q^2f^2(qx)=\mu.
 \label{rs} \ee
Consider the initial value problem for the equation (\ref{rs}):
\bb
f(0)=\tilde a_0<\infty.
\lab{in}
\ee
We fix the initial condition at the point $x=0$ because it
is a fixed point of scaling $x \rightarrow qx$
(there is another such point $x=\infty$ which was considered in
\ci{nov}). The parameter $\mu$ is redundant -- it may be
removed by rescaling of the coordinate and $f(x)$, but we prefer to keep
it as a unique dimensional parameter.

A more general class of the self-similar potentials arises after the
following $q$-periodic closure of the dressing chain \cite{sp2}:
\bb
f_{j+N}(x)=qf_j(qx), \qquad \mu_{j+N}=q^2\mu_j.
\lab{ss2}
\ee
Here we limit our discussion to the $N=1$ case which coincides with
\re{ss1}.

Let $f(z,q)$ denotes a function of two complex variables
that satisfies the equation (\ref{rs}).
Plugging into (\ref{rs}) the power series expansion
\bb
f(z,q)=\sum_{n=0}^{\infty}\tilde{a}_n(q)z^n,
\label{ser}\ee
 one obtains the recursion formula for the coefficients $\tilde{a}_n$:
 \bb (1+q^{n+2})\tilde{a}_{n+1}=\frac{q^{n+2}-1}{n+1}\sum_{s=0}^{n}
\tilde{a}_s\tilde{a}_{n-s}, \qquad n\geq 1,
\label {rec}\ee
$$\tilde{a}_1(1+q^2)=\mu+(q^2-1)\tilde{a}_0^2.   $$
When $f(0,q)=0$, all $\tilde{a}_k$ with even $k$ vanish, i.e. the
$f(z,q)$ is an odd function of $z$ which corresponds to
symmetric potentials.%
\footnote{At least for $|q|<1$, as it will be seen.}
For this case it is convenient to rewrite formulae (\ref{ser}), (\ref{rec})
as follows:
\bb f(z,q)=\sum_{n=1}^{\infty}a_n(q)z^{2n-1},
\label{serie} \ee
\bb
(1+q^{2n})a_n=\frac{q^{2n}-1}{2n-1}\sum_{s=1}^{n-1}a_s
a_{n-s},\qquad a_1=\frac{\mu}{1+q^2}.
\label{recodd}\ee

{ \it Lemma.}\,\, Series (\ref{serie}), (\ref{recodd}) converges 
in the disc
 $|z|<R_q$, for every fixed value of $q$, $|q|<1$.
For the radius of
convergence $R_q$ the following estimate holds:
$R_q\geq \frac{\pi}{2\sqrt{|a_1|\alpha}},$
where $\alpha\equiv\frac{1+|q|^2}{1-|q|^2}$.

 {\it Proof.}
For any natural number $p$ and $|q|<1$ one has:
$|\frac{1-q^{2p}}{1+q^{2p}}|\leq \alpha$. As a result,
$|a_n|\leq c_n$, where $c_n$-coefficients are defined by the same recursion
relations  (\ref{recodd}) with $c_1=|a_1|$ and
the $\frac{q^{2n}-1}{q^{2n}+1}$-factor being
replaced by the $\alpha$ for all $n$.
But the series $\sum_{n=1}^\infty c_n |z|^{2n-1}$ is proportional
to a scaled tangent function of $|z|$.  Hence, our series is majorized by
$$|f(z,q)| \leq
\sqrt{\frac{|a_1|}{\alpha}}\tan{\sqrt{|a_1|\alpha}\,|z|}$$
which asserts the statement of the lemma.
At $q=0$ this gives an exact answer for $R_q$. Using analogous
method one can prove convergence of the series (\ref{ser}) near $z=0$ as well.
\hfill{$\Box$}
\vspace{3mm}

Solutions of (\ref{rs}) for $|q|>1$ can be obtained from those for $|q|<1$
by the following relation: $f(z,q^{-1})=\imath q f(-\imath qz,q)$; therefore
it is sufficient to consider $q$ from the unit disk: $|q|\leq1$.

\smallskip
{\bf Theorem.} \rm For any $q$ such that $|q|<1$, or $q^{2k+1}=1,\,
k=0,1,2,\dots $
in some neighbourhood of $z=0$ there exists the unique solution
$f(z,q)$ of (\ref{rs}) satisfying the initial condition  (\ref{in}).
When $q$ is a primitive root of unity of even degree, $q^{2k}=1$,
existence of analytic solutions depends on the
value of $\tilde a_0$; depending on  $k$
the solution satisfying properly chosen initial condition
may be non-unique.

{\it Proof.} \rm Consider separately the following cases:
\begin{enumerate}
\item Suppose that $|q|<1$. Then, by the given
 $f(0,q)$ one can determine uniquely $f'(0,q),$ $f''(0,q)$, etc by  taking
successive derivatives of (\ref{rs}). Therefore the formal series
(\ref{ser}) is defined  uniquely, and by the  lemma it
converges near the $z=0$ point. So, solution exists and it is unique.%
\footnote{For real $q$, $0<q<1$, this statement follows from a
 general theorem on existence and uniqueness of solutions of differential
 equations with deviating argument of the neutral type proved by
G.A. Kamenskii \cite{EN}.}

\item  Let $q$ be a primitive root of unity of odd degree: $q^{2k+1}=1$.
 In this case recursion relations
(\ref{rec}), (\ref{recodd}) define  uniquely series coefficients. Their
convergence in some neighborhood of $z=0$ follows from the
lemma because the lower bound for $R_q$ can be estimated by replacing
$\alpha\rightarrow\alpha_k$, where $\alpha_k$ is defined by the
inequality $|\frac{1-q^{p}}{1+q^{p}}|\leq \alpha_k$ for any natural $p$ and
$q=\exp{\frac{2\pi \imath}{2k+1}}$. Since $q^n\neq -1$ such finite
$\alpha_k$ does exist but its value depends on $k$.

\item   Let $q$ be a primitive root of unity of even degree:  $q^{2k}=1$.
 Now the left hand side
of the recursion formula (\ref{rec})
 vanishes at $n+2=k(2l+1),\, l=0,1,\dots$, but the right hand
 side does so
only for special values of the parameter $\tilde{a}_0$.
For these specific initial conditions the coefficients
$\tilde{a}_{k(2l+1)-1}$  of the series (\ref{ser}) can be
defined in a non-unique way (i.e. some, or all of them may take
arbitrary values), any other choice of $\tilde{a}_0$
leads to the mismatch in recursion relations for $\tilde{a}_n$
and therefore (\ref{rs}) has no solutions  analytical at $z=0$.
 For example, when $q=-1, \, \mu\neq 0, $ the analytic solution is
unique and exists only for $\tilde a_0=0$ (it is $f(z,-1)=\mu z/2$).
If $\mu=0$ then there is no restriction on $\tilde a_0$ and any even
function $f(z)=f(-z)$ is a solution; the recursion formula (\ref{rec})
 gives in this case an ambiguity in each $\tilde{a}_{2n}$.
 When $q^2=-1$, an analytic solution exists iff $2\tilde{a}_0^2=\mu$.
Then, the derivative $f'(0,q)$ is not determined by the equation (\ref{rs})
and it can take any value. In particular, one has  $f(z,\pm\imath)_{
\tilde a_0=\mu=0}=\tilde a_1 z+\tilde a_5 z^5+\tilde a_9 z^9+\dots,$
where $\tilde a_1, \tilde a_5, \dots$ are arbitrary coefficients such that
the series converges. In section 5 below we give an exact dependence
of the solutions on an arbitrary function for this case.
Similarly, when $q^3=-1$ one should put $\tilde a_0=0$, or
$\tilde a_0^2=\frac{\mu}{1-q^2}$ and  ambiguous
coefficients are $\tilde a_2,\, \tilde a_8,\,\tilde a_{14},\dots $.
\hfill{$\Box$}
\end{enumerate}

\smallskip

Let us make two remarks. First, for the general system of equations
(\ref{dds}), (\ref{dc9}) with $N=2n$ the
uniqueness of its solutions may be violated if
$I_0=2\sum_{i=1}^{N}f_i =0$ (see \cite{vs}).
This condition is obviously satisfied in the self-similar case when $q^N=1$,
but there are also other situations exhibiting non-uniqueness. Second,
analysis of the existence of analytical solutions $f(z,q)$ when
$q=e^{2\pi\imath\phi}$,
$\phi$ -- an irrational number, is essentially a number theory
problem. The nature of irrationality of $\phi$ plays a
crucial role   in the determination of the asymptotic growth of the series
coefficients, in this respect our problem  is close to the question on
the convergence of $q$-hypergeometric series at such $q$.

Since at $|q|<1$ the function $f(z,q)$ is analytic in the neighbourhood
of $z=0$, one can in principle construct $f(z,q)$ on the whole complex
plane as follows. Let us rewrite equation \re{rs} in the form
$f'(z)+f^2(z)=v(qz)$, where $v(z)=q^2(f^2(z)-f'(z))+\mu$, and fix some
number $R$, $0<R<R_q$.  In the
coordinate region $R\leq |z|\leq R/|q|$ this is just the Riccati equation
because $v(qz)$ is the fixed function determined by the values
$f(z,q)$ at $|z|\leq R$.
The general solution of this equation is: $f(z,q)=d/dz \ln{(w_1+cw_2)}$, where
$w_{1,2}$ are independent solutions of the auxiliary Schr\"odinger equation
$-w''(z)+v(qz)w(z)=0$. The constant $c$ is fixed by
some boundary condition.
Continuing this procedure iteratively in the rings
$R/|q|^n\leq |z| \leq R/|q|^{n+1}$, we recover $f(z,q)$ on the  open
complex $z$-plane.
Since all singularities of $f(z,q)$ appear from zeros and singularities
of the solutions of a linear differential equation, one can find  their
qualitative structure.

\smallskip
{\it Proposition 1.}  The function $f(z,q)$ is meromorphic
in the open complex $z$-plane for any fixed $q$, $|q|<1$.

{\it Proof.}  Because $v(z)$ is analytic in the disc
$|z| \leq R$, all permitted singularities of $f(z,q)$ in the first
ring $R \leq |z| \leq R/|q|$ are simple poles with the residues $r_1=1$.
These poles originate from simple zeros, $w(z_p)=0$,
of the wave function $w(z)$ fixed  by the boundary condition
$w'/w|_{z=R} = c_0$.
In the ring $R \leq|z| \leq R/|q|$ the function $v(z)$ may contain double poles
at the points $z = z_p/q$ with the residues equal to $r_1(r_1+1)=2$.
There may be also accompanying simple poles,
but they are irrelevant.  In the vicinity of these singularities
solutions of the auxiliary Schr\"odinger equation have the
form $w(z)\propto (z-z_p/q)^r$ where $r$ is found from the indicial
equation $r(r-1)=r_1(r_1+1)$.  This equation has two solutions: $r=-r_1$ and
$r=r_1+1$ corresponding to linearly independent functions $w_{1,2}$.
The boundary condition $f(R/|q|,q) = c_1$ fixes their relevant combination.
Since $w_{1,2}$ may have only singularities (or zeros) of the form
$(z-z_p/q)^r$ and simple zeros, we conclude that the function $f(z,q)$ in the
second ring can have only pole-type singularities with the integer residues
$r_2=r,$ or $1$. Continuing this procedure iteratively from one ring to another
we prove the assertion.
It is convenient to call the poles with the residues equal to $+1$, which
arise due to the simple zeros of $w(z)$ and are located at $z=z_p$,
as the primary ones.
The poles arising from them as the descendents in the points $z=z_p/q^n$
with the residues equal to one of the possible values of $r\in Z,\, r\neq 1,$
then may be called as the secondary ones.
Some of the series of these secondary poles may be truncated
due to the particular sequence of the residues: $1,\dots, n,\,
-n, \dots,\, -1,\, 0.$  \hfill{$\Box$}

The ``method of steps" described above does not work when $|q|=1$.
\smallskip

When $|q|\ll 1$ it is possible to construct the solution of (\ref{rs})
by expansion of $f(z,q)$ into a perturbation
series over $q$. Plugging $f(z,q)=\sum_{i=0}^\infty h_i(z)q^{2i}$,
$f(0,q)=0$, into
(\ref{rs}) and solving the resulting
differential equations for $h_i(z)$, one obtains the
function $f$. For the first non-vanishing correction we have ($\mu=1$):
$$f(z,0)=h_0(z)=\tanh{z},$$
$$ h_1'(z)+2h_1(z)\tanh{z} +1=0, \qquad h_1(0)=0,$$
$$ f(z,q)=\tanh{z}-q^2\frac{z+\frac{1}{2}\sinh{2z}}{2\cosh^2{z}}+\dots\; .$$

As it was shown above, the properties of the self-similar potentials
differ for $q$ being a root of unity of odd and even degree.
We study below the first
nontrivial cases, $q^3=1$ and $q^4=1$, in detail.

\section{Exact Solution for $\bf q^3=1$}

Let us first derive a general solution of the dressing chain at
$N=3$-periodic closure, which evidently contains as a particular case the
$q^3=1$ self-similar solution. System (\ref{dds}) consisting
of three equations has two invariants of motion generated by
$Tr A=-\lambda I_0-I_1$:
\bb
I_0=2(f_0(x)+f_1(x)+f_2(x)), \label{int}\ee \bb I_1=
\fac13\sum_{i=1}^3 f_i^3 -\sum_{i\neq j}\lambda_i f_j -\fac{1}{24}I_0^3.
\label{integ}
\ee
Deriving explicitly the matrix $A$ one finds
$\beta(x)=\frac{1}{2}I_0f_1+f_0f_2+\lambda_1-\lambda \equiv \gamma(x)-\lambda$.
Substituting $\beta(x)$ in this form into (\ref{sp}) and setting the spectral
parameter $\lambda$ equal to $\gamma(x)$ we obtain:
\bb
-\fac14(\gamma')^2  = (\gamma-E_0)(\gamma- E_1)(\gamma-E_2).
\label{pi}     \ee
This is the differential equation for a Weierstrass $\wp$-function
\cite{abr}:
\bb (\wp')^2 = 4\wp^3 -g_2\wp -g_3, \label{zzzz}\ee
$$\wp(x+2\omega)=\wp(x+2\omega')=\wp(x),\qquad
\wp(x)={1\over x^2 }+ {g_2\over 20}x^2+{g_3\over 28}x^4 +\dots,$$
where $\omega$ and $\omega'$ denote complex semiperiods.

The expression for the potential $u_0$ is obtained by setting equal to zero an
expression in front of $\lambda^2$ in (\ref{sp}):
\bb
u_0(x)=-2\gamma(x)+\sum_{i=0}^2 E_i=2\wp(x+x_0)+\fac13\sum_{i=0}^2E_i.
\lab{potential}
\ee

In the self-similar case one has $I_0=0$ independently on the value of
$\tilde a_0=f_0(0)$. Without loss of generality we can set
$\tilde a_0=0$, which fixes $I_1=0$, i.e. $Tr A=0$.
Constants $E_i$ are just  equal to $\lambda_i$:
$E_i=\lambda_i$. Setting $\lambda_j=\frac{\mu}{q^2-1}
 q^{2j},\,$ and choosing $\mu=q-1$ we get real $u_0(x)$ for real $x$:
\bb
u_0(x)=f_0^2-f_0'+\lambda_0=2\wp(x+\omega_2),\qquad  u_0(0)=-2\lambda_1=2,
\lab{difeq}
\ee
where $\wp(x)$ is the equianharmonic Weierstrass function
and $\omega_2$ is the corresponding real semiperiod, $\omega_2=\omega+\omega'$.
This $\wp$-function is defined by the conditions $g_2=0,\, g_3=4$
and it has definite ratio of semiperiods leading to very simple
transformation rules under the scaling of its argument by $q$:
\bb
\omega'=e^{2\pi\imath/3}\omega, \qquad\Rightarrow \qquad
\wp(qx)=q\wp(x). \label{trp} \ee

The $f_i(x)$ can be determined by solving either the differential
equation (\ref{difeq}), or the system of three algebraic equations
(\ref{int}), (\ref{integ}), and (\ref{potential}):
\bb f_{i}(x)=
-\,\frac{\wp'(x+\omega_{i-1})}{2(\wp(x+\omega_{i-1})+\lambda_i)},
\label{fff}\ee
where we put $\omega_1\equiv q\omega_2,\,\omega_3\equiv q^2\omega_2,\,
\omega_{i+3}\equiv\omega_i.$

The potential $u_0(x)$ is periodic with minima at the points $x_l=2l\omega_2$
and double poles at the points $x_l=(2l+1)\omega_2,\; l\in Z$.
The other two potentials are: $u_1(x)=2\wp(x+\omega_3),\,
u_2(x)=2\wp(x+\omega_1).$
Note that all $f_i$ are complex and only $u_0$ is real. In the Appendix
we present exact solution of the spectral problem for $u_0$ defined
by the zero boundary conditions at singular points.

\section{The Case $\bf q^4=1$ }

 When $q^4=1,\,q=\pm \imath$, the general solution of the equation
 (\ref{rs}) looks as follows:
 \bb 2f(z)=g(z)+\frac{\mu}{g(z)}-\frac{g'(z)}{g(z)},
\label{solu}\ee
where $g(z)$ is an arbitrary function subjected to simple constraints:
\bb g(qz)g(z)=\mu q,\qquad g(0)=(1+q)f(0).  \label{con}  \ee
The conditions (\ref{con}) for $g(z)$  are compatible when
 $f(0)=\pm\sqrt{\mu q}/(1+q)$, i.e. a solution exists only for a
definite value of $f(0)$. In particular, there is no
 function $g(z)$ such that $g(0)=0$ when $\mu\neq0$.

The family of functions $g(z)$ which satisfy the functional equation
(\ref{con})
is rather rich. We consider here two of the possible representations.
The first one is:
\bb g(z)=\pm\sqrt{\mu q}\exp\{F(z^2)\}, \label{first} \ee
where $F(y)$
is an arbitrary odd function of its argument ($y\equiv z^2$).
Corresponding potentials in the Schr\"odinger equation (\ref{dc1}) are
real when $\sqrt{\mu q}$ and $F$ are real:
\bb u(z)=\frac{\partial F(y)}{\partial y}+y\left(\frac{\partial F(y)}{\partial
y}\right)^2 +2y\frac{\partial^2F}{\partial y^2}+
\frac{\mu q}{2}\sinh{2F(y)}
\mp 2\sqrt{\mu qy}\frac{\partial F}{\partial y} e^{F(y)}. \label{ppp}\ee
When $F(y)\rightarrow+\infty$ at $z\to\pm\infty$ and $\mu\neq0$,
potentials (\ref{ppp})
 grow exponentially and, hence, have a purely discrete spectrum. In the
limiting
case $\mu\rightarrow 0$ the choice $F(y)\propto y$
yields the  potential of a harmonic oscillator.

The second solution is related to the (pseudo)lemniscatic Weierstrass
function \ci{abr} which corresponds to the invariants $g_3=0, \, g_2=\pm 1$.
Explicitly,
\bb
f(z)=-\,{1\over 2}{\wp^\prime(z+\frac{q-1}{2}\zeta)-\wp^\prime(\zeta)\over
\wp(z+\frac{q-1}{2} \zeta)-\wp(\zeta)},\qquad
\mu=2\wp(\zeta),
\lab{lem} \ee
\bb {\wp^\prime}^2=4\wp^3\pm \wp,\qquad \omega'=\imath\omega, \qquad
\wp(\pm \imath z)=-\wp(z), \label{trp1} \ee
$$u(z)=2\wp(z+z_0),\qquad z_0=\fac{q-1}{2}\zeta.$$
Taking $z,\, z_0$ to be real we obtain again  the simplest
Lam\'e equation with the real potential.
The corresponding spectral problem is exactly solvable (see Appendix).

Note that for a general $\wp$-function and arbitrary $q$ one has
$\wp(qz;q\omega,q\omega')=q^{-2}\wp(z;\omega,\omega')$. The  $\wp$-function
constructed by the periods $q\omega, q\omega'$ is equal to that
built up by $\omega, \omega'$ if the lattices of periods
$\Gamma=\{\omega,\omega'\}$ and $\Gamma'=\{ q\omega,q\omega'\}$ are
equivalent. But it is known that for two-dimensional lattices the only possible
groups of rotations preserving the lattice are $C_n$ with $n=1,2,3,4,6$.
So, the transformation properties (\ref{trp}) and (\ref{trp1})
of the Weierstrass  $\wp$-function obtained for the cases $q^3=1$ and $q^4=1$
can take place in such a form only for one more non-trivial case when $q^6=1$.
It is possible to extend the analysis to the higher roots of unity
when the potentials are given by \re{its}, using the appropriate
transformation properties of the  Riemann $\Theta$-function.

\section{Realization of the $\bf q$-Oscillator Algebra}

Group-theoretical content of the self-similar
potentials discussed in the previous sections is described by the
$q$-deformed Heisenberg-Weyl, or $q$-oscillator algebra \cite{qosc}:
\bb
a^- a^+ - q^2 a^+ a^- = \mu,\qquad [a^\pm, \mu]=0,
\lab{alg} \ee
where $a^\pm$ are $q$-analogs of the creation and annihilation
operators. The explicit realization of (\ref{alg}) utilizes the
scaling operator $T_q$,
\bb
T_q f(z)=f(qz),\qquad T_q \frac{d}{dz}= \frac{1}{q}\frac{d}{dz} T_q,
\lab{dq} \ee
$$T_q T_r=T_{qr},\qquad T^{-1}_q=T_{q^{-1}},\qquad T_1=1,$$
and looks as follows \cite{sp1}:
\bb
a^+=\sqrt{q}(-\frac{d}{dz}+f(z,q))T_q,\qquad
a^-=\frac{1}{\sqrt{q}}T^{-1}_q(\frac{d}{dz}+f(z,q)),
\lab{cran} \ee
where $f(z,q)$ is a solution for the equation (\ref{rs}) at $x=z/q$.
For simplicity we keep in this section $\mu> 0$.

When $q^2\to 1$ the relations \re{alg} are formally reduced to the
standard bosonic oscillator commutation relations. As it was shown above,
our system has a well-defined limit $q\to 1$ for arbitrary initial condition
(\ref{in}); however, the $q\to -1$ limit exists only when $f(0,q)=0$.
Although in both cases we get a harmonic oscillator potential, the
second case corresponds to the non-standard realization of the
Heisenberg-Weyl algebra:
\bb
b^+=\pm\,\imath (-\frac{d}{dz}+\frac{1}{2}\mu z)P,
\qquad b^-=\mp\,\imath P(\frac{d}{dz}+\frac{1}{2}\mu z),\qquad
[b^-,b^+]=\mu,
\lab{heis} \ee
where $P$ is the parity operator, $Pf(z)=f(-z),\, P^2=1$.
This observation  displays the ability of
the $q$-deformation procedure  to connect continuously
different realizations of the original undeformed algebra.

Consider the limit $q\to 0$. Then,  one has formally $a^-a^+=\mu$.
This algebra admits large variety of realizations,
 e.g. $a^-$ may be a $l\times m$ rectangular matrix.
In our case this limit is singular; however,
the relation $a^-a^+-q^2a^+a^-=\mu$ still may be meaningful.
In particular, if we substitute $z=qx$ into \re{cran}, then
$\lim_{q\to 0} q^2 a^+a^-=-d^2/dx^2 \neq 0$ and $\lim_{q\to 0} a^-a^+=\mu
-d^2/dx^2$ (if $z$ is kept finite, then $q\to 0$ corresponds to the limit
$x\to\infty$ which is not analyzed  in this paper).
Although in this case the self-similar potential is reduced to the known
exactly-solvable one,
we exclude the point $q=0$ from the further discussion because
operators \re{cran} are not well-suited for the algebraic treatment
of corresponding systems.

In the $q$-deformed case the notion of number operator  is not universal.
In order to show this, let
us consider the following combination of the  ladder operators $a^\pm$:
\ba
L&=&a^+a^- -\nu
\lab{hamil} \\
&=&-\, \frac{d^2}{dz^2}+f^2(z)-f'(z)-\nu,\qquad \nu=\frac{\mu}{1-q^2},
  \nonumber \ea
where we assume that $q^2\neq 1$. Operator $L$ satisfies the relations
\bb
L a^\pm =q^{\pm 2} a^{\pm} L,
\lab{ladder}\ee
which  form a $q$-analog  of the spectrum generating algebra
of a harmonic oscillator problem, $[N,b^\pm]=\pm b^\pm,\, N=b^+b^-$.
However, equations \re{ladder} do not imply the existence
of a well-defined number operator $N$ with the properties
$[N,a^\pm]=\pm a^\pm$. Indeed, the algebra \re{alg}, \re{ladder} has
finite-dimensional representations
for which any power of $a^\pm$
does not vanish (see below). The latter means that the spectrum
of $N$ would be unbounded neither from below nor from above, which in turn
would contradict the finite-dimensionality of these cyclic representations.
 Note that usually the relations \re{ladder} are used for the definition of
the formal operator $L$ itself. Sometimes the $q$-oscillator algebra is
defined as a set of identities involving $a^\pm, \mu, L$, and
the inverse $L^{-1}$ as well. We do not assume invertability of $L$, and,
in a sense, utilize the minimal version of the $q$-oscillator algebra
(\ref{alg}), when
the operator $L$ is expressed through $a^\pm$ as it is given in \re{hamil}.

Let us discuss briefly the representation theory of \re{alg},
(\ref{ladder}) paying most attention to the unitarity
and finite-dimensionality of modules.
Denote by $\psi_\lambda^{(r)}$ the eigenstates of the abstract operator $L$,
\bb
L \psi_\lambda^{(r)} =\lambda \psi_\lambda^{(r)},\qquad r=1,\dots, d,
\lab{liegen} \ee
where $\lambda$ is some (complex) number. On the algebraic level nothing
can be said about the degree of degeneracy $d$. For our explicit realization
(\ref{liegen}) is a standard Schr\"odinger equation which has two independent
solutions, i.e. $d=2$. We suppose that the states $\psi_\lambda^{(r)}$ are
uniquely fixed by some (boundary) conditions for all $\lambda$.
Relations \re{ladder} allow to write
\bb
a^+ \psi_\lambda^{(r)}=\alpha_{rs}^+(\lambda q^2)\,\psi_{\lambda q^2}^{(s)},
\qquad
a^- \psi_\lambda^{(r)}=\alpha_{rs}^-(\lambda )\,\psi_{\lambda q^{-2}}^{(s)}.
\lab{action} \ee
Substituting this into \re{alg} and using definition \re{hamil} we deduce
\bb
\alpha_{rs}^\pm(\lambda)=\alpha_r^\pm(\lambda)M_{rs}^{\pm 1}(\lambda), \qquad
\alpha_r^+(\lambda) \alpha_r^-(\lambda)=\nu+\lambda,
\lab{matrix} \ee
where $M_{rs}$ is some non-degenerate matrix. The exact
form of $M_{rs}(\lambda)$ and
$\alpha_r^\pm(\lambda)$ depends on the definition of the basis
vectors $\psi_\lambda^{(r)}$ and can not be found algebraically.

\smallskip
{\it Proposition 2.} There are only five types of
essentilaly different from
each other unitary modules of the algebra \re{alg} ($q\neq 0$) arising at:
1. $\lambda<0$, $0<q^2\leq1$; 2. $\lambda>0$, $0<q^2<1$;
3. $\lambda=0$, $q\neq \pm1$;  4. $\lambda \neq 0$, $-1<q^2<0$;
5. $\lambda \neq 0$, $q^2=-1$.

{\it Proof.} We call the module unitary when the operators $a^\pm$ are
hermitian conjugates of each other in the corresponding basis of states.
The hermiticity conditions are ensured by $\lambda$,
$q^2$ being real  and the
choice $\alpha^+_r(\lambda)=(\alpha^-_r(\lambda))^*=\sqrt{\nu+\lambda},\,
M^\dagger=M^{-1}$.
Let us consider first the interval  $0<q^2\leq 1$.
For negative eigenvalues $\lambda$ the representation should be
truncated from below; otherwise $\nu +\lambda $ will start to be negative.
So, we get the highest-weight infinite-dimensional representation:
\ba
L|j\rangle&=&-\,\nu q^{2j}|j\rangle, \nonumber   \\
a^+|j\rangle&=&\sqrt{\nu(1-q^{2(j+1)})}|j+1\rangle, \qquad j=0,1,\dots
\nonumber   \\  a^-|l\rangle&=&\sqrt{\nu(1-q^{2l})}|l-1\rangle,
\qquad\qquad l=1,2,\dots
\nonumber   \\ a^-|0\rangle&=&0, \lab{1d} \ea
which is generated by the vacuum state $|0\rangle$ corresponding
to the particularly chosen $\psi_{-\nu}^{(r)}$. Note that we
absorbed the matrix $M_{rs}$ \re{matrix} into the definition of
states $|j\rangle$. In our case vacuum state is unique because
it is defined by the first-order differential equation.
The eigenvalues $\lambda_j=-\nu q^{2j}$ provide the physical
discrete spectrum of the infinite-soliton system described by our
self-similar potential at real $z$, $0<q^2<1$, $\mu>0$ \cite{sh,sp1}.
The module \re{1d} is well-defined when $q^2\to 1$.

For positive $\lambda$ we get a non-highest-weight infinite-dimensional
unitary module
\ba L|j\rangle_{\lambda}&=&\lambda q^{2j}|j\rangle_{\lambda},
\qquad\qquad\qquad j\in Z \nonumber  \\
a^+|j\rangle_{\lambda}&=&\sqrt{\nu+\lambda q^{2(j+1)}}|j+1\rangle_{\lambda},
\nonumber  \\  a^-|j\rangle_{\lambda}&=&
\sqrt{\nu+ \lambda q^{2j}}|j-1\rangle_{\lambda}.
\lab{2c} \ea
Here we assume that the module is generated by some particularly chosen
eigenstate $|0\rangle_\lambda$ of the Hamiltonian $L$ with eigenvalue
$\lambda$, and we  absorbed  again the  matrix $M_{rs}$ into the definition
of states $|j\rangle_{\lambda}$. Note that in this case $a^+$ lowers the
energy. For our specific realization of the $q$-oscillator algebra
wave functions $|j\rangle_{\lambda}$ form a particular subset of the
scattering states of the infinite-solitonic potential.
The limit $q^2\to 1$ is not defined for \re{2c}.

A peculiar situation is described by the
subspace of states corresponding to zero eigenvalue of $L$.
 In this case $a^+$ and $a^-$ represent the
integrals of motion, $[a^+,a^-]=[L,a^\pm]=0$,
and they can be diagonalized simultaneously with $L$:
\bb L\psi^{(r)}=0,\qquad
a^\pm\psi^{(r)}=\sqrt{\nu}\, e^{\pm\imath \theta_r}\, \psi^{(r)}.
\lab{3c} \ee
This is a $d$-dimensional (for any real $q^2, \, q\neq \pm1$) reducible
representations. The irreducible one-dimensional components are the simplest
cyclic representations for which the number operator can not
be defined. In our case $d=2$ and in principle all angles $\theta_r$
may be found explicitly.
The wave functions $\psi^{(r)}$ provide the simplest coherent states
of the $q$-oscillator algebra.

Finally, when $-1\leq q^2 <0$ we have the following possibilities.
At $|q|<1$ the infinite-dimensional representation with the defining relations
as in (\ref{1d}) arises, but  now the successive eigenvalues of $L$
have different signs. 
At $q=\pm\imath$ we have the following finite-dimensional representation:
\ba
L|0\rangle_\lambda=\lambda|0\rangle_\lambda, \quad\qquad\qquad
L|1\rangle_\lambda=-\,\lambda|1\rangle_\lambda,~~~~~
\nonumber \\
a^+|0\rangle_\lambda=\sqrt{\nu-\lambda}|1\rangle_\lambda, \qquad
a^+|1\rangle_\lambda=\sqrt{\nu+\lambda}|0\rangle_\lambda,
\nonumber \\
a^-|0\rangle_\lambda=\sqrt{\nu+\lambda}|0\rangle_\lambda, \qquad
a^-|1\rangle_\lambda=\sqrt{\nu-\lambda}|0\rangle_\lambda,
\lab{fermi} \ea
where $\lambda$ belongs to the interval $[-\nu, \nu]$.
In  this case the $q$-oscillator algebra is reduced to the following
superalgebra:
\bb
\{ a^-, a^+ \} = \mu, \quad \{ a^\pm , a^\pm \} = \mu_\pm, \quad
[\mu, a^\pm ] = [\mu_\pm ,a^\mp ]= [\mu_\pm,\mu]=0.
\lab{super} \ee
The operators $\mu_\pm$ may be set equal to zero, which
corresponds to the boundary values of the parameter $\lambda$
in (\ref{fermi}), $\lambda=\pm\nu$, and this gives
the standard fermionic oscillator algebra.
\hfill{$\Box$} \medskip

Discussion of the physical significance of the states \re{3c}, and
analysis of the structure of modules provided by the self-similar
potentials at $-1\leq q^2< 0$ lies beyond the scope of the present work.

It is easy to see that the representations can be finite-dimensional
iff $\lambda(q^n-1)=0,\, q \neq\pm 1.$ Indeed,
the ordinary bosonic oscillator algebra arising at $q=\pm 1$
does not have finite-dimensional representations; the possibility $q=0$  was
already excluded due to the exoticity. The case $\lambda=0$
was considered above and it corresponds to the $d$-dimensional,
in general reducible, representation. At $\lambda\neq 0$
all modules are constructed from series of states $\psi^{(r)}_{\lambda q^{2j}}$
for some range of $j$. Let $q$ be not a root of unity. Then the
set $\{\psi^{(r)}_{\lambda q^{2j}}\}$ can be finite-dimensional iff
it is truncated
from below and above due to the zeros of $\alpha^\pm_{rs}(\lambda)$.
The equation $a^-\psi^{(r)}_\lambda=0$ holds at $\lambda=-\nu$ and
$a^+\psi^{(r)}_{\lambda'}=0$ at $\lambda'=-\nu q^{-2}$. Evidently, these states
belong to two different irreducible infinite-dimensional highest-weight
representations. We thus conclude that $q^n=1,\, q^2\neq 1$ is a necessary
condition for the existence of finite-dimensional representations at
$\lambda\neq 0$.

Finite-dimensionality of the modules
at $q^n=1$ ($n$ is the lowest number satisfying this identity)
emerges due to the existence of the non-trivial central elements of the
algebra:
\bb
C^\pm=(\pm \imath a^\pm)^{\lbrack\!\lbrack n \rbrack\!\rbrack},
\qquad [C^\pm, a^\mp]=[C^\pm,L]=0,
\lab{center} \ee
where ${\lbrack\!\lbrack n \rbrack\!\rbrack}= n$ for odd $n$ and
${\lbrack\!\lbrack n \rbrack\!\rbrack} =n/2$ for even $n$.
Due to the definition of $L$ the states $\psi_{\lambda q^{2j}},\,
j=0,\dots,{\lbrack\!\lbrack n \rbrack\!\rbrack} $
form the ${\lbrack\!\lbrack n \rbrack\!\rbrack}$-dimensional
irreducible module which
can not be unitary at $q^2\neq  -1$.
The non-zero $C^\pm$ correspond to the cyclic representations
(``cyclic" oscillators) for which the number operator does not exist.
By purely algebraic means one finds that $C^\pm$ satisfy  the
following polynomial algebra:
\bb
C^\pm C^\mp = \nu^{\lbrack\!\lbrack n \rbrack\!\rbrack} -
(-L)^{\lbrack\!\lbrack n \rbrack\!\rbrack},
\lab{polynomalg} \ee
which is the particular case of algebras, considered
in \cite{shyam,vs} from a different point of view. A subset of states,
for which the right hand side of \re{polynomalg} vanishes, form the highest
weight finite-dimensional representation. Note that the operator
$L^{\lbrack\!\lbrack n \rbrack\!\rbrack}$ enters \re{polynomalg}
with different signs for $n=4k$ and $n=4k+2$.

Returning to our explicit realization of the $q$-oscillator
algebra we note that for odd $n$ the $C^\pm$ are ordinary differential
operators of the $n$-th order whose commutativity with the
Hamiltonian defines the particular cases of hyperelliptic potentials
\re{its}. In general the operators $C^\pm$ are not equal -- this depends on the
choice of the integrals of motion. E.g., the $q^3=1$ system \re{fff}
gives $C^+=C^-\equiv C$,
\bb
C=\imath(\frac{d^3}{dz^3} -\fac32\{\wp, \frac{d}{dz}\}), \qquad
C^2=L^3+1.
\lab{example} \ee
For the spectral problem considered in the Appendix, the $L^3+1$ operator
is self-adjoint, but it is easy to see that
its square root $C$ does not preserve imposed boundary conditions.
One cannot, however, exclude the possibility that for some of the spectral
problems the operators $C^\pm$ represent real physical observables.

\bigskip\bigskip
\noindent
{\Large{\bf Acknowledgements}}
\bigskip

The authors are grateful to A.Shabat for a general guidance and
participation in this research. We would like also to thank V.Kac and
L.Vinet for stimulating discussions.

\bigskip \noindent
{\Large \bf Appendix }
\bigskip

Though the spectrum of a particle described by the Schr\"odinger equation
\bb       -\psi''+2\wp(x)\psi=\lambda\psi, \qquad
 \psi(0)=\psi(2\omega_2)=0,
 \label{schr}\ee
 is known
(see, for example \cite{op}), we find it to be useful to show here how it can
be
derived from the chain (\ref{dds}). This derivation serves as an application
of the factorization method to the  spectral problems
associated with singular potentials
given by the (hyper)elliptic functions which appear after the periodic
closure of the dressing chain.

It was found that $F(x)= \frac{\pm w-\beta'(x)}{2\beta(x)}$ is the solution
of the Riccati equation (\ref{spec}), related to the Schr\"odinger
equation (\ref{schr}). Taking $x=0$ to be a singular point of $\beta(x)$ and
plugging in $\beta(x)=-\wp(x)-\lambda$, we obtain:
\bb -F(x)=\frac{\psi'}{\psi}=\frac{\pm w+\wp'(x)}{2(\wp(x)+\lambda)},
\label{yyyy}\ee
where $w^2=(trA)^2-4detA=-4(\lambda^3+1)$ for the $q^3=1$ system.
Integrating (\ref{yyyy}), one obtains the general solution of
(\ref{schr}):
\ba \psi(x)&=& c_1\exp{\left(-\int\frac{w-\wp'}{2(\wp+\lambda)}dt\right)}+
c_2\exp{\left(\int
\frac{w+\wp'}{2(\wp+\lambda)}dt\right)} \nonumber \\
&=&c \sqrt{\wp+\lambda}
\sin{\left(\sqrt{\lambda^3+1}\int_{0}^{x}\frac{dt}{\wp(t)+
\lambda}+\delta\right)}. \ea
The  quantization
condition  $\psi(0)=0$ yields $\delta=0$. When $x\to 0$,
$\sqrt{\wp+\lambda}\sim x^{-1} $,
but $\sin{(\int_{0}^{x}\frac{dt}{\wp(t)+\lambda})}\sim x^3$,
 so that eigenfunction $\psi(x)\sim x^2$ as
$x\rightarrow 0$.
The condition $\psi(2\omega_2)=0$
yields the transcendental equation for $\lambda$ which determines the spectrum:
\bb
\sqrt{\lambda^3+1}\int_{0}^{\omega_2}
\frac{dt}{\wp(t)+\lambda} =\fac12 \pi n.
\label{trans}
\ee
The $\lambda\rightarrow\infty$ asymptotics is:
$$\lambda^{(n)}=\frac{\pi^2n^2}{4\omega_2^2},\qquad
\omega_2=\fac12 \int_1^{+\infty}\frac{dx}{\sqrt{x^3-1}}=
\fac16 B(\fac12,\fac16 )  \approx 1.21 .$$
This is the spectrum of a particle in the infinitely deep well,
as might be expected.  For the first three
eigenvalues numerical solution gives:
$\lambda^{(1)}=2.34 ,\, \lambda^{(2)}=5.39, \,   \lambda^{(3)}=9.18$.
For any $\lambda^{(n)}$ one can find that $\psi \sim
(x-2\omega_2)^2$ when $x\rightarrow 2\omega_2$.

This method of derivation of spectrum is not restricted to the
equianharmonic Weierstrass
function (self-similar case); the straightforward generalization of
 (\ref{trans}) yields the spectrum
for arbitrary $\wp$-function (\ref{zzzz}):%
\footnote{The choice of constants
$E_i$ is restricted to the case when the corresponding
potential is real and has singularities over the real axis.}
\bb
 \prod_{i=0}^2 \sqrt{\lambda-E_i}\, \int_{0}^{\omega_2}
\frac{dx}{\wp(x)+\lambda} = \fac12 \pi n, \qquad
\prod_{i=0}^2(\lambda-E_i)=\lambda^3-\fac14(g_2\lambda-g_3).
\lab{last} \ee

The procedure presented above is not applicable in general for the $q^4=1$
potentials, because in this case the matrix element $a_{12}$ of the product
of four Darboux transformations vanishes identically on the
solutions of (\ref{dds}). However, the spectra of the
particular cases corresponding to the (pseudo)lemniscatic
Weierstrass function can be found from the formula \re{last}.

The obtained purely discrete spectrum (\ref{trans}) originates from the
requirement of the finiteness of wave functions  in the  singular points
of the potential.
It is possible to remove singularities from the real axis by shifting
the argument of the $\wp$-function
along the imaginary axis. This leads in general to a complex potential.
In particular, one cannot get a real potential without singularities
in the case of the equianharmonic and pseudolemniscatic $\wp$-functions
(when two of the roots $E_i$ of the polynomial $4x^3-g_2x-g_3$ are complex).
For the lemniscatic Weierstrass function the shift of the coordinate
$z\rightarrow z+\omega'$ leads to the real periodic bounded potential
$(-\fac12 <\wp<0)$ with one finite gap in the spectrum.
So, for the latter case one has two physically different self-adjoint
spectral problems.

\newpage

\bebib{30}

\bi{infeld} Infeld, L. and Hull, T.D., {\it Rev.Mod.Phys.} {\bf 23}, 21
(1951).

\bi{sh} Shabat, A., {\it Inverse Prob.}  {\bf 8}, 303 (1992).

\bi{sp1} Spiridonov, V., {\it Phys.Rev.Lett.} {\bf 69}, 398 (1992).

\bi{sp2} Spiridonov, V., in {\it Proc. of the XIX$^{th}$ ICGTMP}, Salamanca,
29 June - 4 July 1992, to appear; Preprint UdeM-LPN-TH-123, 1992,
Comm.Theor.Phys. (Allahabad), to appear.

\bi{qa} Kulish, P.P. and Reshetikhin, N.Yu., {\it Zap. Nauchn. Sem. LOMI}
Vol. {\bf 101} (Nauka, 1981) p.101 (in Russian);
Sklyanin, E.K., {\it Funct.Anal.Appl.} {\bf  17}, 273 (1983);
Drinfel'd, V.G., in {\it Proc. of the ICM}, Berkeley, 1986;
Jimbo, M., {\it Lett.Math.Phys.} {\bf 11}, 247 (1986).

\bi{qosc} Biedenharn, L.C., {\it J.Phys.}  {\bf A22}, L873 (1989);
Macfarlane, A.J., {\it J.Phys.}  {\bf A22}, 4581 (1989);
Hayashi, T., {\it Comm.Math.Phys.} {\bf 127}, 129 (1990).

\bi{qsf} Exton, H., {\it $q$-Hypergeometric Functions and Applications},
Ellis Horwood, Chichester, 1983; Atakishiyev, N.M. and Suslov, S.K., in
{\it Progress in Approximation Theory}, A.A.Gonchar and E.B.Saff (eds.),
Springer, 1991.

\bi{qpol} Vaksman, L.L. and Soibel'man, Ya.S., {\it  Funct.Anal.Appl.}
{\bf  22}, 170 (1988); Masuda, T., Mimachi, K., Nakagami, Y.,
Noumi, M., Saburi, Y., and Ueno, K., {\it Lett.Math.Phys.} {\bf 19},
187, 195 (1990); Floreanini, R. and Vinet, L., {\it Lett.Math.Phys.} {\bf 22},
45 (1991); Kalnins, E.G., Manocha, H.L., and Miller, W., {\it J.Math.Phys.}
{\bf 33}, 2365 (1992).

\bi{root}
Roche, P. and Arnaudon, D., {\it Lett.Math.Phys.} {\bf 17}, 295 (1989);
De Concini, C. and Kac, V.G., {\it Progr.Math.} {\bf 92}, 471 (1990).

\bi{shyam} Shabat, A.B. and Yamilov, R.I., {\it Leningrad Math.J.}
        {\bf 2}, 377 (1991).

\bi{its} Its, A.R. and Matveev, V.B., {\it Sov.J.Theor.Math.Phys.}
{\bf 23}, 343 (1975).

\bi{vs} Shabat, A.B. and Veselov, A.P., Preprint of Forschungsinstitut f\"ur
Mathematik ETH Z\"urich, 1992,
to appear in {\it Funct.Anal.Appl.}

\bibitem{EN} El'sgol'ts, L.E. and Norkin, S.B., {\it Introduction to the Theory
and Application of Differential Equations with Deviating Arguments},
Academic Press, 1973;
Kamenskii, G.A., {\it Mat.Sbornik} {\bf 55(97)}, 363 (1961).

\bi{nov} Novokshenov, V.Yu., Clarkson University preprint, INS-203, 1992.

\bi{abr} Abramowitz, M. and Stegun, I.A., {\it Handbook of
Mathematical Functions}, Nat. Bur. of Stand., 1966.

\bibitem{op} Olshanetsky, M.A. and Perelomov, A.M., {\it Phys.Rept.}
{\bf 94}, 322 (1983).

\end{thebibliography}
\end{document}